\documentclass[12pt]{article}
\usepackage{bezier}
\usepackage{amssymb}
\usepackage{epsfig}

\newcommand{\nc}{\newcommand}
\nc{\be}{\begin{equation}}
\nc{\ee}{\end{equation}}
\nc{\bea}{\begin{eqnarray}}
\nc{\eea}{\end{eqnarray}}

\nc{\eqn}[1]{{(\ref{#1})}}
\nc{\cA}{{\cal A}}
\nc{\cB}{{\cal B}}
\nc{\cC}{{\cal C}}
\nc{\cD}{{\cal D}}
\nc{\cE}{{\cal E}}
\nc{\cF}{{\cal F}}
\nc{\cG}{{\cal G}}
\nc{\cH}{{\cal H}}
\nc{\cI}{{\cal I}}
\nc{\cJ}{{\cal J}}
\nc{\cK}{{\cal K}}
\nc{\cL}{{\cal L}}
\nc{\cM}{{\cal M}}
\nc{\cN}{{\cal N}}
\nc{\cO}{{\cal O}}
\nc{\cP}{{\cal P}}
\nc{\cQ}{{\cal Q}}
\nc{\cR}{{\cal R}}
\nc{\cS}{{\cal S}}
\nc{\cT}{{\cal T}}
\nc{\cU}{{\cal U}}
\nc{\cV}{{\cal V}}
\nc{\cW}{{\cal W}}
\nc{\cX}{{\cal X}}
\nc{\cY}{{\cal Y}}
\nc{\cZ}{{\cal Z}}
\nc{\simo}[1]{{\stackrel{#1}{\simeq}}}
\nc{\geqo}[1]{{\stackrel{#1}{\geq}}}
\nc{\geo}[1]{{\stackrel{#1}{>}}}
\nc{\guo}[1]{{\stackrel{#1}{\succ}}}

\nc{\rbo}{\raisebox}
\nc{\RR} {\rangle \! \rangle}
\nc{\LL} {\langle \! \langle}
\nc{\rmi}[1]{{\mbox{\small #1}}}
\nc{\eq}{eq.~}
\nc{\nr}[1]{(\ref{#1})}
\nc{\ul}{\underline}
\nc{\mc}{\multicolumn}
\nc{\todo}[1]{\par\noindent{\bf $\rightarrow$ #1}}

\nc{\cu}{{\cal u}}

\title{
  \begin{flushright} {\small $\begin{array}{ l } 
\mbox{SPhT-00/013} \\
\mbox{HD--THEP--00--09} \\
    \end{array} $}
 \end{flushright}
Chiral symmetry restoration and
axial vector renormalization
for Wilson fermions}

\author{T.~Reisz$^{a,b}\,$\thanks{Supported by a Heisenberg Fellowship}
        $\;$ and
       H.~J.~Rothe$^{b}$,
         \\ \\$^a$ Service de Physique Th\'eorique de Saclay \\
        CE-Saclay,\\
        F-91191 Gif-sur Yvette Cedex, France
         \\ \\$^b$ Institut
        f\"ur Theoretische Physik,\\
        Universit\"at Heidelberg, \\
        Philosophenweg 16, \\
        D-69120 Heidelberg, Germany}

\begin{document}

\maketitle

\begin{abstract}
Lattice gauge theories with Wilson fermions break chiral symmetry. 
In the $U(1)$ axial vector current this manifests itself 
in the anomaly. On the 
other hand it is generally expected that the axial vector 
flavour mixing current is non-anomalous. We give a short, 
but strict proof of this to all orders of perturbation theory, 
and show that chiral symmetry restauration implies a 
unique multiplicative renormalization constant for the current. This 
constant is determined entirely from an irrelevant operator 
in the Ward identity. The basic ingredients 
going into the proof are the lattice Ward identity, 
charge conjugation symmetry and the power counting theorem. 
We compute the renormalization constant to one loop order. 
It is largely independent of the particular lattice realization 
of the current.
\end{abstract}

%
%
%
%
\section{Introduction}

Any realization of fermions on the lattice has to
respect the constraints imposed by the
Nielsen-Ninomiya theorem \cite{nielsen_ninomiya}.
Whereas Wilson fermions break chiral symmetry explicitly,
Ginsparg-Wilson fermions \cite{ginsparg_wilson, neuberger}
have an exact chiral symmetry on
the lattice that is generated by composite local lattice operators
\cite{luescher1}.
In both cases, the continuum chiral flavor mixing symmetry 
and the anomaly are to be properly reproduced as the cutoff
is removed.

In a recent paper \cite{anomaly1} it was shown that under 
very general conditions
on the lattice Dirac operator, which, in particular, are satisfied both for
Wilson and for Ginsparg-Wilson fermions,
the axial anomaly is correctly generated in the continuum limit.
The main conditions are gauge invariance, absense of doublers, and
locality on the lattice in a more general sense.
The origin of the anomaly is traced back to an irrelevant, local
lattice operator in the axial vector Ward identity.

For Ginsparg-Wilson fermions, the composite operator
in the chiral transformation which ensures  
an exact flavor mixing symmetry on the lattice,
stays irrelevant under renormalization \cite{gw_renorm}.
As a consequence, the axial vector current does not require
renormalization.
On the other hand, for Wilson fermions it is not obvious,
although widely believed, that the chiral symmetry
becomes restored in the continuum limit. 
Below we give a short but strict proof of this assertion
to all orders of perturbation theory, 
based on lattice power counting for massless theories \cite{massless_pct}. 
Although we explicitely refer to Wilson fermions, the result is as 
general as that for the anomaly generation given in \cite{anomaly1}. 
As we shall show, the only role played by the irrelevant, symmetry
breaking operator in the flavor mixing axial vector 
Ward identity is to give rise to a unique
multiplicative renormalization $Z_j$ of the axial vector current, 
ensuring that chiral symmetry is restored in the continuum limit.
We compute the one-loop contribution to $Z_j$ 
as a function of the Wilson parameter. 
The result is largely independent of the particular lattice 
regularization of the current.

\section{General framework}

Although our general proof will be given for QED, it generalizes in an 
obvious way to non-abelian gauge theories with massless fermions. 

\subsection{Renormalized lattice QED}

The action for renormalized QED is given by
\be
  S(A, \psi, \overline\psi) \; = \;
      S_W(U) + S_f(U,\psi,\overline\psi) + S_{gf}(A) .
\ee
$S_W(U)$ is e.g.~the Wilson plaquette action
\be
   S_W(U) = Z_A \, \frac{1}{2g^2} \sum_{x\in a{\mathbb Z}^4} 
    \sum_{\mu\not=\nu=0}^{3}
    \biggl( 1 -  U(x;\mu) U(x+a\widehat\mu;\nu)
      U(x+a\widehat\nu;\mu)^{-1} U(x;\nu)^{-1} 
    \biggr) ,
\ee
with $g$ the renormalized gauge coupling constant and
$U(x,\mu)=\exp{( iagA_\mu(x))}$ $\in$ $U(1)$.
The fermion action is given by
\be \label{flm.ferm}
  S_{f} \; = \; a^4 \sum_{x\in a {\mathbb Z}^4} Z_\psi
   \overline{\psi}(x) \left( D[U] + m_0 \right) \psi(x) ,
\ee
with $\psi$ a 2-flavor Dirac spinor field and with
$D[U]$ the Wilson Dirac operator,
\bea
   D[U]\psi(x) 
   & = & \frac{1}{2a} \sum_{\mu=0}^3 
   \biggl\lbrack (\gamma_\mu-r) U(x;\mu) \psi(x+a\widehat\mu)
   \nonumber \\
   && \qquad - (\gamma_\mu+r) 
       U(x-a\widehat\mu;\mu)^{-1} \psi(x-a\widehat\mu)
    + 2r \psi(x)
   \biggr\rbrack .
\eea
$m_0$ is the bare fermion mass,
which for massless fields must be tuned to its critical value
of $O(g^2)$.
$S_{gf}$ denotes the gauge fixing action.
For concreteness we choose the Lorentz gauge,
\be \label{flm.gf_lorentz}
   S_{gf}(A) \; = \; a^4 \sum_{x\in a{\mathbb Z}^4} \;
     \frac{\lambda}{2} \;
     \left( \sum_{\mu=0}^3 \frac{1}{a} 
            \widehat{\partial}_\mu^* A_\mu(x) 
     \right)^2 ,
\ee
with $\lambda>0$ the gauge fixing parameter.
Here and in the following, $\widehat\partial_\mu$ and
$\widehat{\partial}_\mu^*$ denote the forward and backward 
lattice difference operators, respectively,
\be
  \widehat\partial_\mu f(x) = f(x+a\widehat\mu) - f(x),\;
   \widehat\partial_\mu^* f(x) = f(x) - f(x-a\widehat\mu),
\ee
where $\widehat\mu$ is the unit vector in $\mu$ direction.

The generating functional
$W$ of the connected correlation functions is given by
\bea \label{flm.conn_gen_functl}
   && \exp W(J,\eta,\overline\eta) \; = \; \int
    \prod_x \left( d\psi(x) d\overline\psi(x) \prod_\mu dA_\mu(x)
            \right) \;
   \nonumber \\
   && \quad \cdot \quad \exp
    \biggl\lbrace - S(A,\psi,\overline\psi)
           + S_c(A,\psi,\overline\psi;J,\eta,\overline\eta)
    \biggr\rbrace 
\eea
with source term $S_c$
\bea
   && S_c(A,\psi,\overline\psi;J,\eta,\overline\eta) = a^4 \sum_x 
    \biggl\lbrace \sum_\mu J_\mu(x) A_\mu(x) 
      + \overline\eta(x)\psi(x) + \overline\psi(x)\eta(x)
    \biggr\rbrace .
\eea
The vertex functional $\Gamma$ is obtained by a
Legendre transformation
\be \label{flm.LeTr_bare}
   W(J,\eta, \overline\eta) = \Gamma(\cA,\psi,\overline\psi)
    + a^4 \sum_x 
    \biggl( \sum_\mu J_\mu(x) \cA_\mu(x) + \overline\eta(x)\psi(x)
      + \overline\psi(x)\eta(x)
    \biggr),
\ee
where
\be
   a^4 \cA_\mu(x) = \frac{\partial W}{\partial J_\mu(x)}, \quad
   a^4 \psi(x) = \frac{\partial W}{\partial\overline\eta(x)}, \quad
   a^4 \overline\psi(x) = - \frac{\partial W}{\partial\eta(x)} .
\ee
By $\widetilde\Gamma^{(n,m)}(k,l)$
we denote the momentum space
vertex function of $n$ fermion pairs and $m$ gauge fields,
with their collected momenta denoted by $k$ and $l$,
respectively.
For $Q$ any composite local lattice operator, we write 
$\widetilde\Gamma^{(n,m)}_Q(q;k,l)$
for the vertex function
with one insertion of $Q$, with $q$ its momentum.
Momentum conservation is implied.
Massless fermions require that
\be \label{flm.zero_mass}
   {\rm tr\,} \widetilde\Gamma^{(1,0)}(k=0) \; = \; 0
\ee
to be achieved by tuning $m_0$,
where the trace is taken in spinor space.
$Z_A$ and $Z_\psi$ are uniquely determined by appropriate normalization
conditions at non-exceptional momenta, e.g.~by
\be \label{flm.ren_cond}
   \biggl.
     \frac{i}{4} \frac{\partial}{\partial\widetilde{k}_0}
     {\rm tr\,} \gamma_0 \widetilde\Gamma^{(1,0)}(k)
   \biggr\vert_{\overline{k}} \; = \; 1 ,
   \quad
   \biggl.
     - \frac{1}{2} \frac{\partial}{\partial\widehat{k}_0}
      \widetilde\Gamma^{(0,2)}_{11}(k)
   \biggr\vert_{\overline{k}} \; = \; \widehat{\overline{\mu}} ,
\ee
where $\overline{k}=(\overline\mu\not=0,0,0,0)$,
$\widehat{k}=(2/a)\sin(ka/2)$,
$\widetilde{k}=(1/a)\sin(ka)$.

\subsection{Symmetries}

Below we make explicit reference to the
charge conjugation symmetry
\be \label{flm.cgt_symm}
   \Gamma(\cA^C,\psi^C,\overline\psi^C) \; = \;
     \Gamma(\cA,\psi,\overline\psi) ,
\ee
where
\be \label{flm.cgt}
   \cA_\mu^C(x) = - \cA_\mu(x) , \;
   \psi^C(x) = C \overline\psi(x)^T , \;
   \overline\psi^C(x) = -\psi(x)^T C^{-1} .
\ee
The superscript $T$ denotes transposition and
$C$ the charge conjugation matrix satisfying
\be
   C^{-1} \gamma_\mu C \; = \; - \gamma_\mu^T ,
    \quad \mu=0,\dots,3 .
\ee
Furthermore, applying a gauge transformation leads to the
local Ward identity
\bea  \label{flm.WI_gauge_bare}
   &&  i \sum_{\mu=0}^3 \frac{1}{a} \widehat\partial_\mu^* 
    \frac{\partial\Gamma}{\partial\cA_\mu(x)}
    + \biggl\lbrack g \overline\psi(x) 
         \frac{\partial\Gamma}{\partial\overline\psi(x)}
         - g \psi(x) \frac{\partial\Gamma}{\partial\psi(x)}
      \biggr\rbrack
   \nonumber \\
   && \quad  - i \lambda a \sum_{\mu,\nu=0}^3 
         \widehat\partial_\nu^*\widehat\partial_\nu
         \widehat\partial_\mu^* \cA_\mu(x)
    \; = \; 0 .
\eea
It implies that the renormalized action is of the form
as stated above.
\section{Chiral symmetry breaking and symmetry restoration}

Chiral symmetry is broken by the Wilson Dirac operator.
Under a local, flavor mixing chiral transformation
\be
   \delta\psi(x) = i \epsilon(x) \; \sigma_\alpha \gamma_5 \psi(x) ,
    \quad
    \delta\overline\psi(x) = i \epsilon(x) \; 
     \overline\psi(x)
     \gamma_5 \sigma_\alpha ,
\ee
where $\sigma_\alpha$, $\alpha=1,2,3$, denote the Pauli matrices
acting in flavour space,
the action transforms according to
\be \label{flm.delta_s}
   \delta S \; = \; a^4 \sum_x i \epsilon(x) 
    \left( - \sum_\mu \frac{1}{a} \widehat{\partial}_\mu^* j_{\mu\alpha}(x)
           \; + \; \Delta_\alpha(x) \; + \; 2 m_0 \cP_\alpha(x)
    \right) ,
\ee
with 
\be
   \cP_\alpha(x) \; = \; Z_\psi \; \overline\psi(x)
     \gamma_5 \sigma_\alpha \psi(x) .
\ee 
The gauge invariant local operators
$j_\mu$ and $\Delta$ are not uniquely determined
by (\ref{flm.delta_s}).
In general, $\Delta$  is a local lattice operator which is
classically irrelevant, that is, 
\be
   \lim_{a\to 0} \Delta_\alpha(x) \; = \; 0 .
\ee
It has UV degree 4 and IR degree 5.
A convenient representation of $j_{\mu\alpha}$ and of $\Delta_\alpha$
is given by
\bea \label{flm.j_delta_reps}
   j_{\mu\alpha}(x) 
   & = & Z_\psi \frac{1}{2} 
       \biggl( \overline\psi(x) (\gamma_\mu + s)
          \gamma_5 \sigma_\alpha U(x;\mu) \psi(x+a\widehat\mu)
   \nonumber \\
   && \qquad + \overline\psi(x+a\widehat\mu) (\gamma_\mu - s)
          \gamma_5 \sigma_\alpha U(x;\mu)^{-1} \psi(x)
       \biggr) ,
   \nonumber \\
   \Delta_\alpha(x) 
   & = & - Z_\psi \frac{1}{2a} \sum_{\mu=0}^3
    \biggl\lbrace (r-s) \;
        \overline\psi(x) \gamma_5 \sigma_\alpha 
   \\
   && \qquad \biggl\lbrack 
          U(x;\mu) \psi(x+a\widehat\mu)
           +  U(x-a\widehat\mu;\mu)^{-1} \psi(x-a\widehat\mu)
           -  2 \psi(x)
             \biggr\rbrack
    \nonumber \\
   && + (r+s) \;
        \biggl\lbrack 
           \overline\psi(x+a\widehat\mu) U(x;\mu)^{-1}
           + \overline\psi(x-a\widehat\mu) U(x-a\widehat\mu;\mu)
           - 2 \overline\psi(x) 
        \biggr\rbrack
   \nonumber \\
   && \qquad 
       \gamma_5 \sigma_\alpha \psi(x)
     \biggr\rbrace ,
   \nonumber
\eea
where $r$ is the Wilson parameter, and $s$ some arbitrary real but
otherwise fixed constant.
The following discussion on renormalization does not depend on
a particular choice of s.

We add to the source part of the action $S_c$ a term
\be \label{flm.composite_source}
   a^4 \sum_{x} \sum_{\alpha=1}^3
     \biggl( \sum_{\mu=0}^3 G_{\mu\alpha}(x) j_{\mu\alpha}(x)
      + F_\alpha(x) 
       \biggl\lbrack \Delta_\alpha(x) + 2 m_0 \cP_\alpha(x) \biggr\rbrack
     \biggr) ,
\ee
and denote the corresponding vertex functional by
$\Gamma^{\,\prime}(\cA,\psi,\overline\psi;G,F)$. Obviously,
$\Gamma^{\,\prime}(\cA,\psi,\overline\psi;G=0,F=0)$ $=$
$\Gamma(A,\psi,\overline\psi)$. 
Then (\ref{flm.delta_s}) implies that $\Gamma^{\,\prime}$
satisfies the axial vector current Ward identity
\bea \label{flm.local_wi}
   && \sum_\mu \frac{1}{a} \widehat\partial_\mu^*
    \frac{\partial \Gamma^{\,\prime}}{\partial a^4 G_{\mu\alpha}(x)}
  \; + \; 
    \biggl\lbrace \frac{\partial \Gamma^{\,\prime}}{\partial a^4 \psi(x)}
        \sigma_\alpha \gamma_5 \psi(x)
       - \overline\psi(x) \sigma_\alpha \gamma_5
          \frac{\partial \Gamma^{\,\prime}}{\partial a^4 \overline\psi(x)}
    \biggr\rbrace 
   \\
   && \qquad\qquad = \; 
      \frac{\partial \Gamma^{\,\prime}}{\partial a^4 F_\alpha(x)} 
      \; + \; O(F,G) .
   \nonumber
\eea

The functional identity (\ref{flm.local_wi}) is equivalent to the
infinite set of momentum space Ward identities
\be \label{flm.corr_wi}
   i \; \sum_{\mu=0}^3 \widehat{q}_\mu 
         \widetilde\Gamma^{(n,m)}_{j_{\mu\alpha}}(q;k,l) 
     \; - \;
     \widetilde\Gamma_{QED}^{(n,m)}(k,l)
     \; = \;
     \widetilde\Gamma^{(n,m)}_{\Delta_\alpha}(q;k,l)
     +
     2 m_0 \; 
     \widetilde\Gamma^{(n,m)}_{\cP_\alpha}(q;k,l).
\ee
Here we have written $\widetilde\Gamma_{QED}^{(n,m)}(k,l)$
for the pure QED part,
which is a linear combination of $\widetilde\Gamma^{(n,m)}(k,l)$
with $\gamma_5\sigma_\alpha$ attached to the various external
fermion lines, but with no composite operator inserted.
According to the renormalization prescription of QED,
it is UV finite and universal in the continuum limit.

QED is already renormalized, but
because of $\Delta\not\equiv 0$, the axial vector current
$j_\mu$ requires additional renormalization.
This renormalization is multiplicative.
That is, there exists a renormalization constant
$Z_j$ such that
\be
    \widetilde\Gamma^{(n,m)}_{j_{\mu\alpha}R}(q;k,l) \; = \; 
    Z_j \widetilde\Gamma^{(n,m)}_{j_{\mu\alpha}}(q;k,l) 
\ee
is finite in the continuum limit, for all $n$ and $m$.
The renormalized current satisfies the Ward identities
\be \label{flm.corr_wi_renorm}
   i \; \sum_{\mu=0}^3 \widehat{q}_\mu 
         \widetilde\Gamma^{(n,m)}_{j_{\mu\alpha} R}(q;k,l) 
     \; - \;
     \widetilde\Gamma_{QED}^{(n,m)}(k,l)
     \; = \;
     \widetilde\Gamma^{(n,m)}_{\Delta_\alpha R}(q;k,l),
\ee
where
\bea \label{flm.delta_ren}
   && \widetilde\Gamma^{(n,m)}_{\Delta_\alpha R}(q;k,l) 
     \; = \; 
     \widetilde\Gamma^{(n,m)}_{\Delta_\alpha }(q;k,l)  
     \nonumber \\
   && \qquad + \; 
     \biggl\lbrack
        \widetilde\Gamma^{(n,m)}_{\cP_\alpha }(q;k,l) +
        i \left( Z_j-1 \right)
        \sum_{\mu=0}^3 \widehat{q}_\mu 
         \widetilde\Gamma^{(n,m)}_{j_{\mu\alpha} }(q;k,l)
     \biggr\rbrack 
\eea
are the renormalized vertex functions with one 
$\Delta_\alpha$-insertion.
Because of 
\be
   m_0 = O(g^2) \quad \mbox{and} \quad 
   Z_j - 1 = O(g^2) ,
\ee
the part in brackets on the right hand side of (\ref{flm.delta_ren})
is equivalently obtained by adding
local counter terms to the lattice source action $S_c$,
that is, in (\ref{flm.composite_source}), the square bracket
is replaced by
\be 
   \Delta_\alpha(x) + 2 m_0 \cP_\alpha(x)
   + (Z_j-1) \frac{1}{a} \sum_{\mu=0}^3
     \widehat{\partial_\mu^*} j_{\mu\alpha}(x) .
\ee
We now show that for the particular choice of $Z_j$,
these counter terms provide precisely overall
Taylor subtractions at zero momentum, 
for all correlation functions with
one $\Delta$-insertion, according to their overall ultaviolet lattice
divergence degrees. 
Together with (\ref{flm.ren_cond}),
because $\Delta$ is an irrelevant local lattice operator,
this then implies that
\be
   \lim_{a\to 0} \widetilde\Gamma^{(n,m)}_{\Delta_\alpha R}(q;k,l)
   \; = \; 0,
\ee
to all orders of perturbation theory, and for all $n$ and $m$
\cite{massless_pct}.
The renormalized axial vector current becomes conserved
in the continuum limit.

For the proof of this assertion, we recall that
$\Delta$ is a local operator of IR degree 5.
This implies that
$\widetilde\Gamma_{\Delta R}^{(1,1)}$ is continuous at zero momentum
and $\widetilde\Gamma_{\Delta R}^{(1,0)}$ is once continuously
differentiable at zero momentum.
(These are the only vertex functions with one $\Delta$-insertion
that require overall UV subtractions, with overall (lattice) divergence
degrees $0$ and $1$, respectively.)

First, charge conjugation symmetry implies that
\be \label{flm.delta_11}
   \widetilde\Gamma^{(1,1)}_{\Delta_\alpha R}(0) \; = \; 0 .
\ee
Furthermore, for the massless theory, satisfying
(\ref{flm.zero_mass}), we obtain from 
(\ref{flm.corr_wi_renorm}) with $n=1$, $m=0$
\be \label{flm.delta_0}
   \widetilde\Gamma^{(1,0)}_{\Delta_\alpha R}(0) \; = \; 0,
\ee
because $\widetilde\Gamma^{(1,0)}_{j_{\mu\alpha}R}(q;k,l)$
is at most logarithmically infrared divergent if all
momenta are sent to zero.
Again, using charge conjugation symmetry,
we know that for small momenta
\be \label{flm.delta_10}
   \widetilde\Gamma^{(1,0)}_{\Delta_\alpha R}(q;k) \; = \;
     i \sum_{\mu=0}^3 \gamma_\mu q_\mu \gamma_5 \sigma_\alpha \;
        + \; o(q,k)
     \quad \mbox{as}\; q,k\to 0 ,
\ee
with (infrared) finite constant $z_\Delta$.
Hence, order by order, $Z_j$ is uniquely determined by 
the requirement that
\be \label{flm.delta_irrelev}
   \widetilde\Gamma^{(1,0)}_{\Delta_\alpha R}(q;k) \; = \;
        o(q,k) \quad\mbox{as}\; q,k\to 0 .
\ee
This completes the proof.

\section{Axial Vector Renormalization Constant in One-Loop order}

To one-loop order the renormalization constant $Z_j$ is determined from the 
condition (\ref{flm.delta_irrelev}), 
where ${\widetilde\Gamma}^{(1,0)}_{\Delta R}$
is given according to (\ref{flm.delta_ren}) by
\be \label{flm.oneloop}
  \biggl\lbrack
    {\widetilde\Gamma}^{(1,0)}_{\Delta_\alpha R}(q;k)
  \biggr\rbrack_{1-loop}
    \; = \; 
      \biggl\lbrack 
         {\widetilde\Gamma}^{(1,0)}_{\Delta_\alpha}(q;k)
      \biggr\rbrack_{1-loop}
    \; + \; 2m_0\gamma_5 \sigma_\alpha
    \; + \;  i \, (Z_j-1) \; 
      \sum_{\mu=0}^3 {\widehat q}_\mu\gamma_\mu\gamma_5 \sigma_\alpha .
\ee
The lattice Feynman diagrams that contribute to the first term on the 
right hand side are listed below. The vertices that correspond
to the $\Delta$-insertion are given in Appendix B.
\bea
   && \nonumber \\
   && 
     \begin{picture}(15.0,10.0)
         \epsfig{bbllx=1237,bblly=333,
                 bburx=2417,bbury=253,
                 file=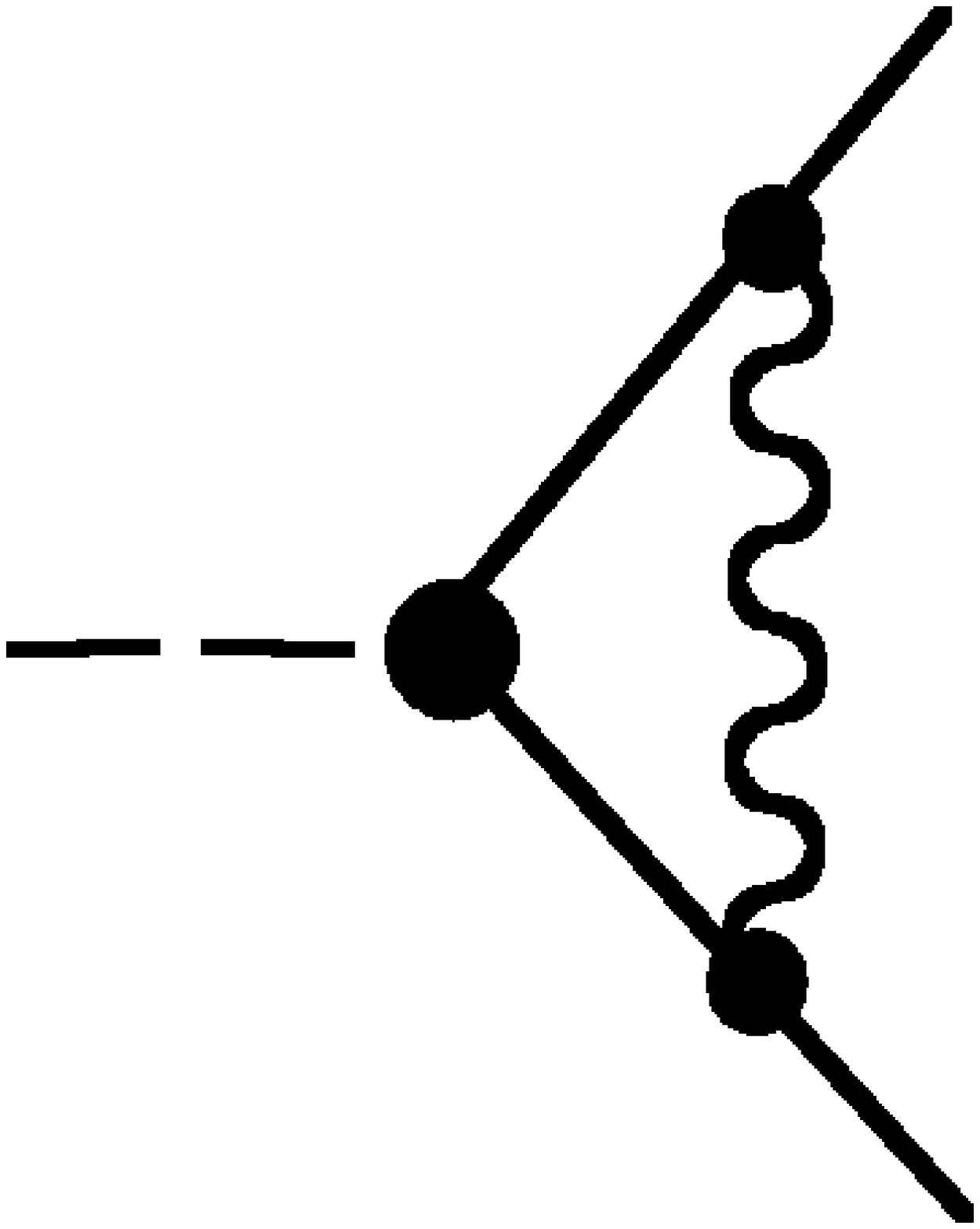,
                 scale=0.12}
     \end{picture}
     \begin{picture}(15.0,10.0)
         \epsfig{bbllx=837,bblly=333,
                 bburx=2017,bbury=253,
                 file=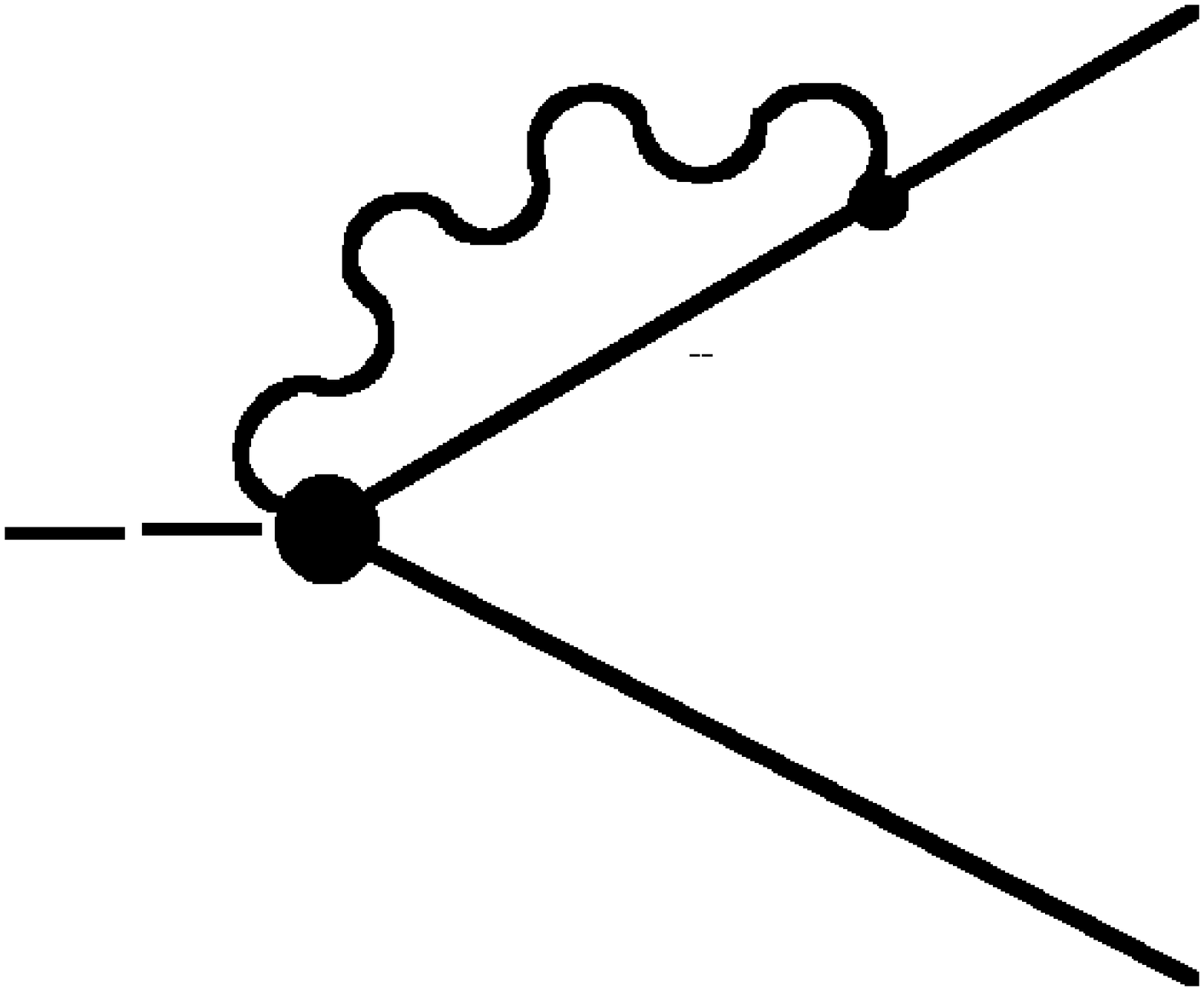,
                 scale=0.10}
     \end{picture} 
     \begin{picture}(15.0,10.0)
         \epsfig{bbllx=337,bblly=333,
                 bburx=1517,bbury=253,
                 file=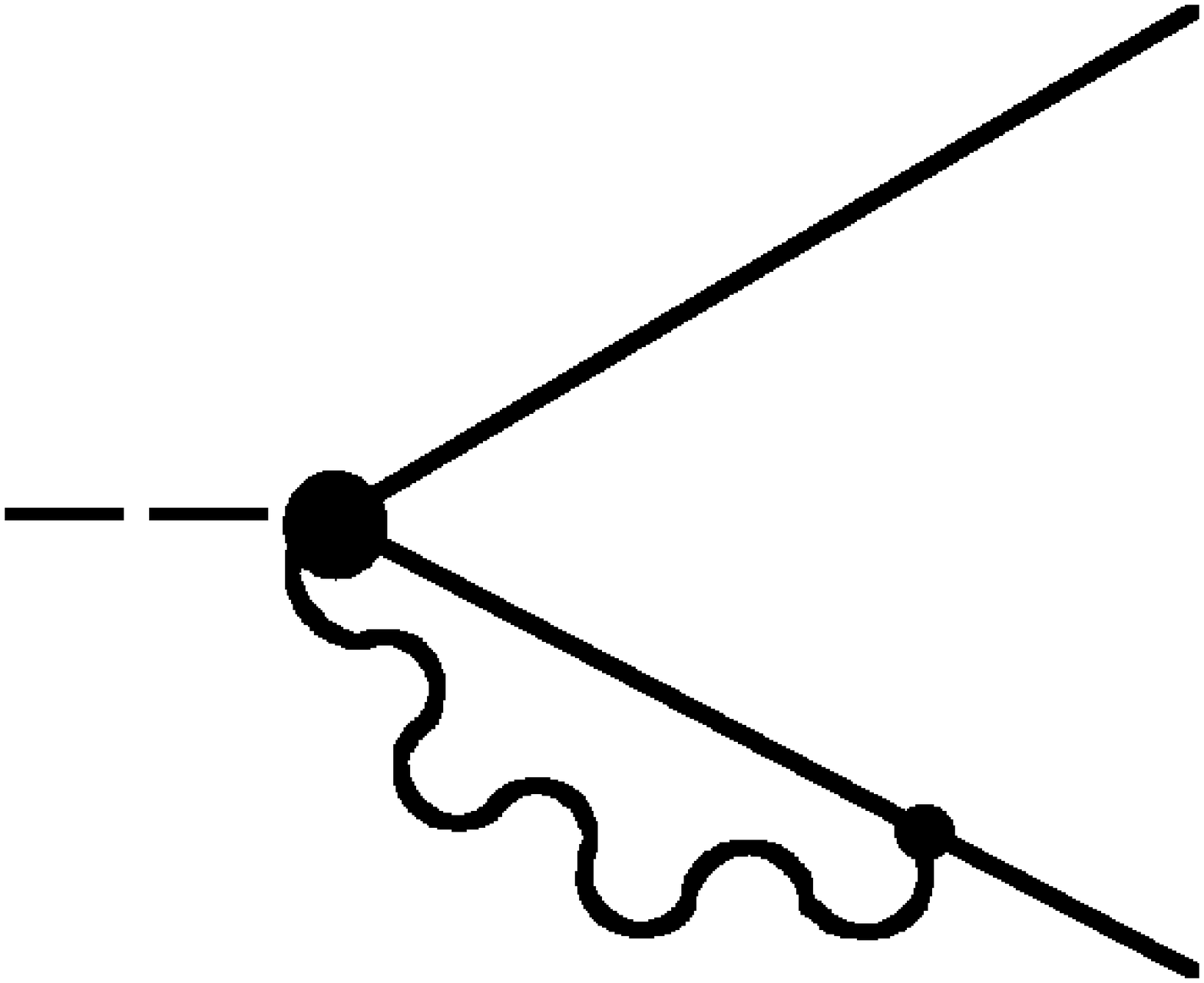,
                 scale=0.10}
     \end{picture} 
     \begin{picture}(15.0,10.0)
         \epsfig{bbllx=-163,bblly=333,
                 bburx=1017,bbury=253,
                 file=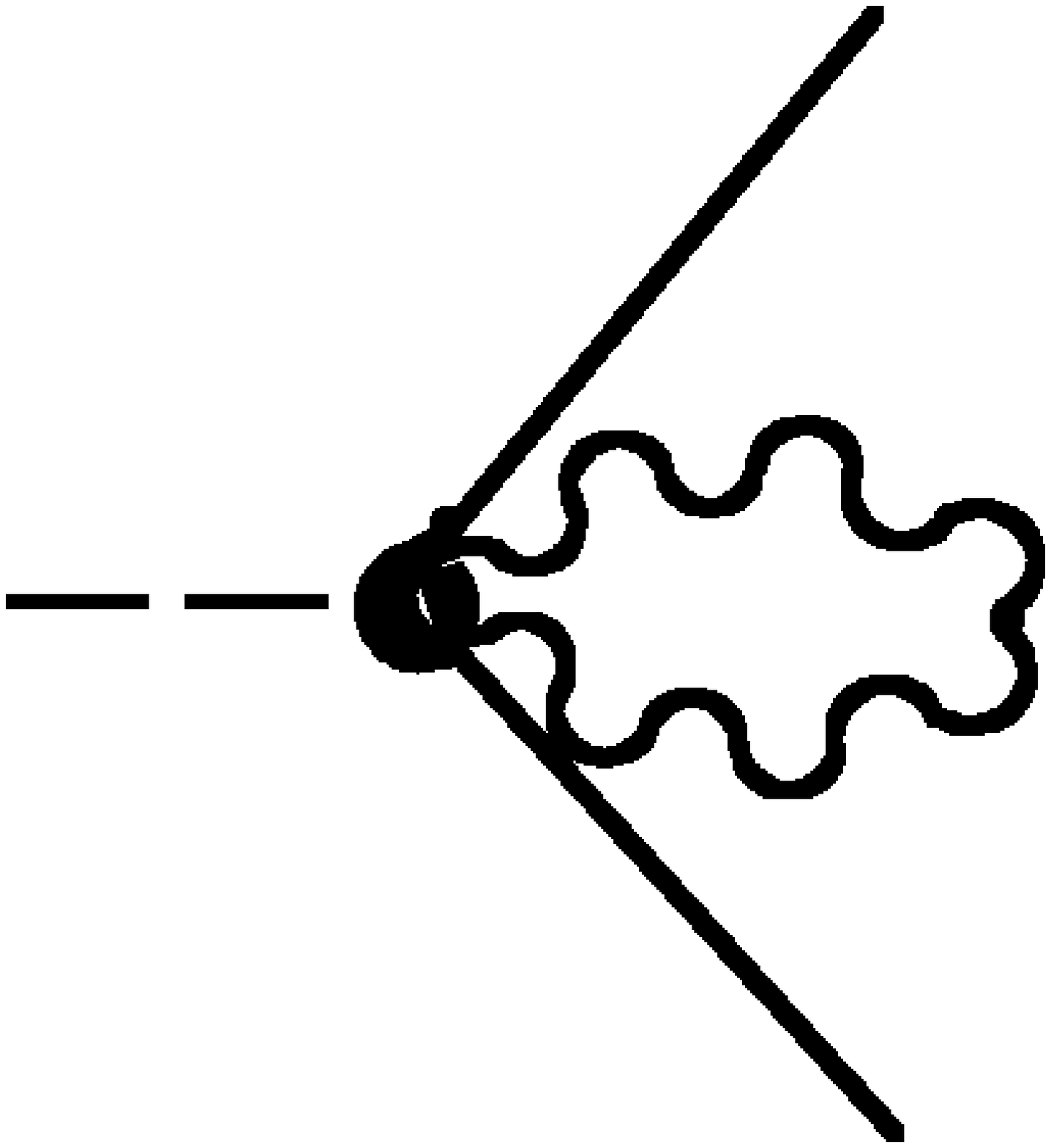,
                 scale=0.12}
     \end{picture} 
   \\
   && \nonumber
\eea
To implement condition (\ref{flm.delta_irrelev}) we expand the 
right hand side of (\ref{flm.oneloop}) around vanishing momentum
up to first order.
After a tedious but straight forward calculation one finds that
\be
   T_1 \; 
     {\widetilde\Gamma}^{(1,0)}_{\Delta_\alpha R}(q;k)
   \; = \; i \, \biggl\lbrack -\zeta_\Delta 
        \; + \; \left( Z_j-1 \right) \biggr\rbrack
   \; \sum_{\mu=0}^3 q_\mu\gamma_\mu\gamma_5 \sigma_\alpha ,
\ee
where $T_1$ denotes the Taylor expansion to order $1$
at zero momentum.
${\zeta}_\Delta$ is given by the following (r-dependent) expression.
\be \label{flm.zdelta_int}
   \zeta_\Delta \; = \; 
     - \; C g^2 r^2 \; 
       \int_{-\pi}^\pi \frac{d^4\ell}{(2\pi)^4} \;
     \frac{h_\sigma (\ell)}{({\widetilde\ell}^2+{\widehat M}^2_r)^2} ,
\ee
with $C=1$ for $U(1)$ and $C=(N^2-1)/(2N)$ for $SU(N)$, and
\bea
  && h_\sigma(\ell) \; = \; \cos\ell_\sigma \;
    \left( \; {\widetilde\ell}^2 - \; {\widetilde\ell}^2_\sigma 
       + \frac{{\widehat\ell}^{\, 2}}{2} \; \eta_\sigma(\ell) \right)
       \; - \; {\widetilde\ell}^2_\sigma \; \eta_\sigma(\ell) 
  \nonumber \\
  && \qquad + \; \frac{{\widetilde\ell}^2}{2} + \frac{1}{4} \; r^2 
       {\widehat\ell}^{\, 2}
    \left( \; {\widehat\ell}^{\, 2} \cos^2\frac{\ell_\sigma}{2}
           - \; {\widetilde\ell}^2_\sigma 
    \right) ,
\eea
where
\bea
   && \widehat{\ell}_\mu \; = \; 2 \sin\frac{\ell_\mu}{2} , \;\;
     \widehat\ell^{\, 2} \; = \; \sum_{\mu=0}^3 \widehat\ell^{\, 2}_\mu , \;
     \widetilde{\ell}_\mu \; = \; \sin\ell_\mu , \;\;
     \widetilde\ell^2 \; = \; \sum_{\mu=0}^3 \widetilde\ell^2_\mu , 
     \nonumber \\
   && \widehat M_r \; = \; \frac{r}{2} \; \widehat\ell^{\, 2} , \quad
     \eta_\sigma(\ell) \; = \; 2 \cos^2\frac{\ell_\sigma}{2}
        \; - \; \sum_{\mu=0}^3 \cos^2\frac{\ell_\mu}{2} .
\eea
$\sigma$ is any one of the indices $0,\dots ,3$.
Condition (34) now determines $Z_j$ to be
\be
     Z_j \; = \; 1 \; + \; {\zeta}_\Delta .
\ee
$\zeta_\Delta$ 
does not depend on a particular realization
of the axial vector current $j_{\mu\alpha}$, 
Eqn.~(\ref{flm.j_delta_reps}), that is, it is independent of
the parameter $s$.

\begin{figure}[h]

\begin{center}
\setlength{\unitlength}{0.8cm}

%
%
\begin{picture}(15.0,10.0)

%
%
%

\epsfig{bbllx=37,bblly=63,
        bburx=1217,bbury=253,
        file=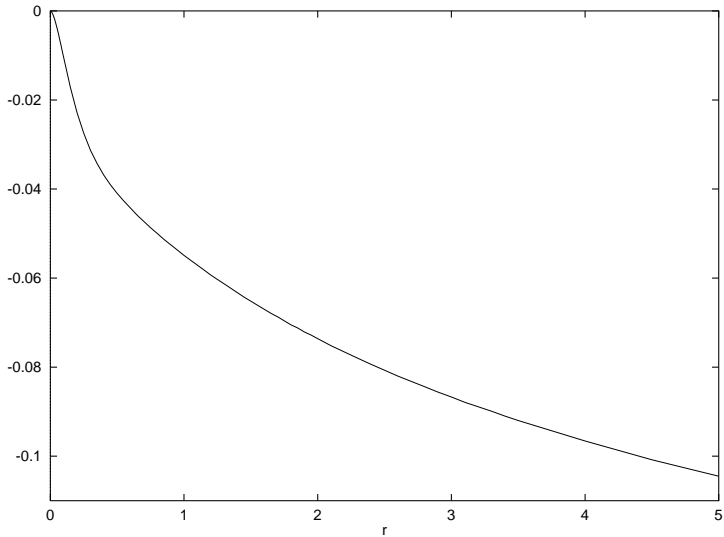,
        scale=0.8}

%
%

\end{picture}
%
%
\end{center}

\caption{\label{flm.pic.zdelta}
$\zeta_\Delta(r)/(Cg^2)$ as function of the Wilson parameters $r$. 
We have $\zeta_\Delta(-r)$ $=$ 
$\zeta_\Delta(r)$ and $\zeta_\Delta(r)\not=0$
whenever $r\not= 0$.
}
\end{figure}

Note that the (finite) renormalization constant of the axial vector current 
is solely determined from diagrams involving the insertion 
of a classically irrelevant operator. 
The integral (\ref{flm.zdelta_int}) is evaluated numerically. 
The dependence of 
$\zeta_\Delta$ on the Wilson parameter is shown in 
Fig.~\ref{flm.pic.zdelta}. From this 
figure we see that $\zeta_\Delta$ is a monotonically decreasing function 
of $r>0$. There exists no non-zero value of $r$ 
for which the axial current is not renormalized.
In particular, for the commonly used value $r=1$ we
obtain
$\zeta_\Delta=-0.0549 C g^2$.

\section{Conclusion}

In this paper we have investigated the renormalization of the 
flavour mixing axial vector current for massless gauge theories 
with Wilson fermions.
The corresponding axial vector Ward identity involves a symmetry
breaking lattice operator $\Delta$ which is local and classically
irrelvant.
Using the lattice power counting theorem for massless
field theories, we have shown that $\Delta$
uniquely determines the renormalization constant of 
the axial vector current in such a way
that the chiral symmetry becomes restored in the continuum limit,
to all orders of perturbation theory.

We have computed the renormalization constant to one-loop order.
It is largely independent of a particular lattice realization
of the current and non-vanishing whenever the Wilson
parameter $r\not=0$.

Although we have considered Wilson fermions, the result of symmetry
restoration in the continuum limit is quite general. 
It holds for any lattice Dirac operator
that satisfies a general set of conditions. These conditions are
gauge invariance and charge conjugation symmetry, 
absense of doublers, and locality in the more general sense as stated in 
\cite{anomaly1,luescher_jansen}.

\begin{appendix}

%
%
\section{\label{app}Small momentum behavior}

The infrared properties of $\Delta$ ensure that the
vertex functions 
$\widetilde\Gamma_{\Delta R}^{(1,1)}(q;k,l)$ 
and $\widetilde\Gamma_{\Delta R}^{(1,0)}(q;k)$
are continuous and once continuously differentiable
at zero momentum, respectively.
Their regular parts are obtained from the small momentum
behavior of that part of the vertex functional $\Gamma^{\,\prime}$
that is linear in the source $F$.
In momentum space it reads
\bea
   && \int \frac{d^4 q}{(2\pi)^4} \frac{d^4 k_1}{(2\pi)^4} 
           \frac{d^4 k_2}{(2\pi)^4} \; \sum_{\mu\alpha}
      \widetilde{F}_{\mu\alpha}(q)
    \biggl\lbrace
       (2\pi)^4 \delta(q+k_1+k_2) 
   \nonumber \\
   && \qquad \widetilde{\overline\psi}(k_1) 
       \biggl\lbrack i (\rho_r k_2+\rho_l k_1)_\mu \gamma_\mu + \eta
       \biggr\rbrack  \gamma_5 \sigma_\alpha
      \widetilde\psi(k_2)
   \\
   && + \xi \widetilde{\overline\psi}(k_1)  
        \gamma_\mu A_\mu(-q-k_1-k_2)\gamma_5 \sigma_\alpha
      \widetilde\psi(k_2)
    \biggr\rbrace
\eea
with c-numbers $\rho_r,\rho_l,\eta$ and $\xi$.
Applying the transformation (\ref{flm.cgt}) yields the same
expression with
\be
   \rho_r \leftrightarrow \rho_l , \;
   \eta \rightarrow \eta, \;
   \xi \rightarrow -\xi .
\ee
The symmetry (\ref{flm.cgt_symm}) thus implies that
$\rho_r=\rho_l$ and $\xi=0$.
The vanishing of $\eta$ is implied by the chiral Ward identity.
This implies the statements 
(\ref{flm.delta_11})-(\ref{flm.delta_10}).

\section{\label{appb}Feynman rules}

We state the Feynman rules for the insertion of one 
$\Delta_\alpha$-operator, Eqn.~(\ref{flm.j_delta_reps}),
that are required for the computation of the axial vector
current renormalization constant $Z_j$ to one-loop order.
For simplicity the rules are given for gauge group U(1).
\bea
   && \nonumber \\
   && \begin{picture}(5.0,1.0)
         \epsfig{bbllx=837,bblly=533,
                 bburx=2017,bbury=403,
                 file=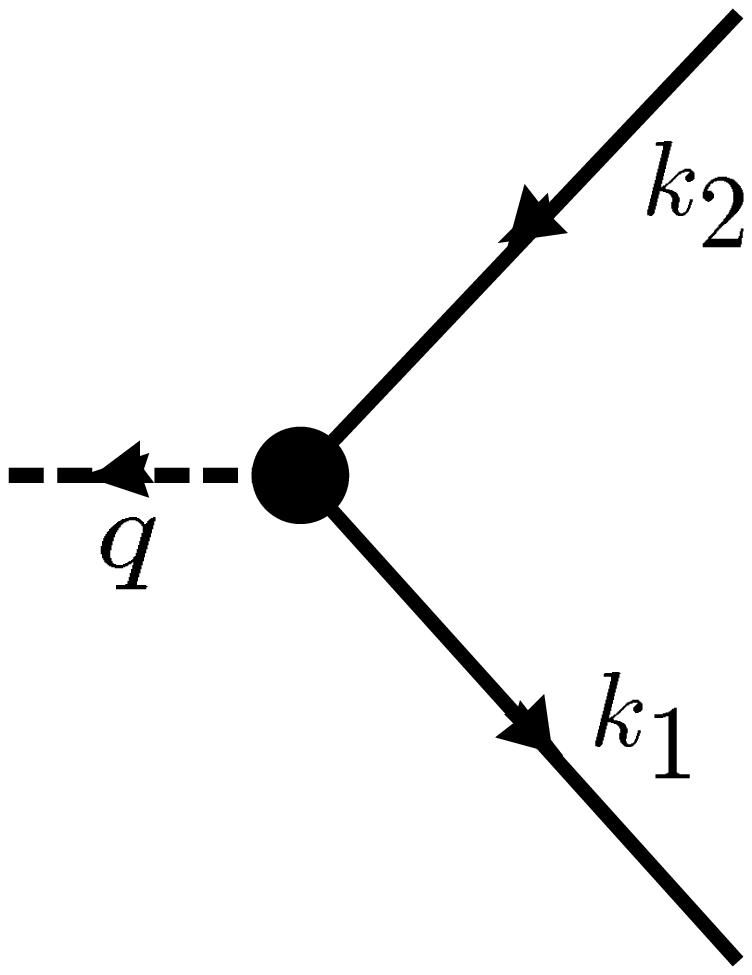,
                 scale=0.25}
    \put(-320.0,0.0){\makebox(1.0,0.0){
      $= \; \frac{a}{2} \; \biggl\lbrack (r-s) \; \widehat{k_2^2}
              + (r+s) \; \widehat{k_1^2} \biggr\rbrack 
        \gamma_5 \sigma_\alpha$}}
      \end{picture}
   \\
   && \nonumber
\eea
\bea
   && \nonumber \\
   && \begin{picture}(5.0,1.0)
         \epsfig{bbllx=837,bblly=533,
                 bburx=2017,bbury=403,
                 file=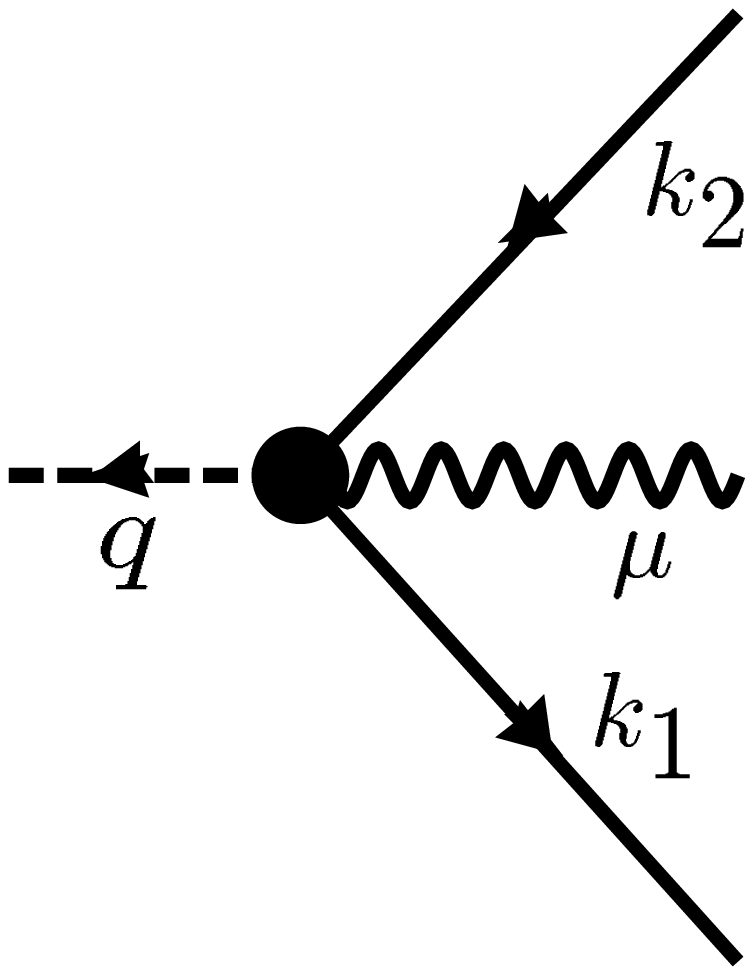,
                 scale=0.25}
    \put(-270.0,0.0){\makebox(1.0,0.0){
      $= \; g\; \frac{a}{2} \; \biggl\lbrack 
          (r-s) \; \widehat{(q+k_1+k_2)}_\mu
              - (r+s) \; \widehat{(q-k_1-k_2)}_\mu \biggr\rbrack 
        \gamma_5 \sigma_\alpha$}}
      \end{picture}
   \\
   && \nonumber
\eea
\bea
   && \nonumber \\
   && \begin{picture}(5.0,1.0)
         \epsfig{bbllx=837,bblly=533,
                 bburx=2017,bbury=403,
                 file=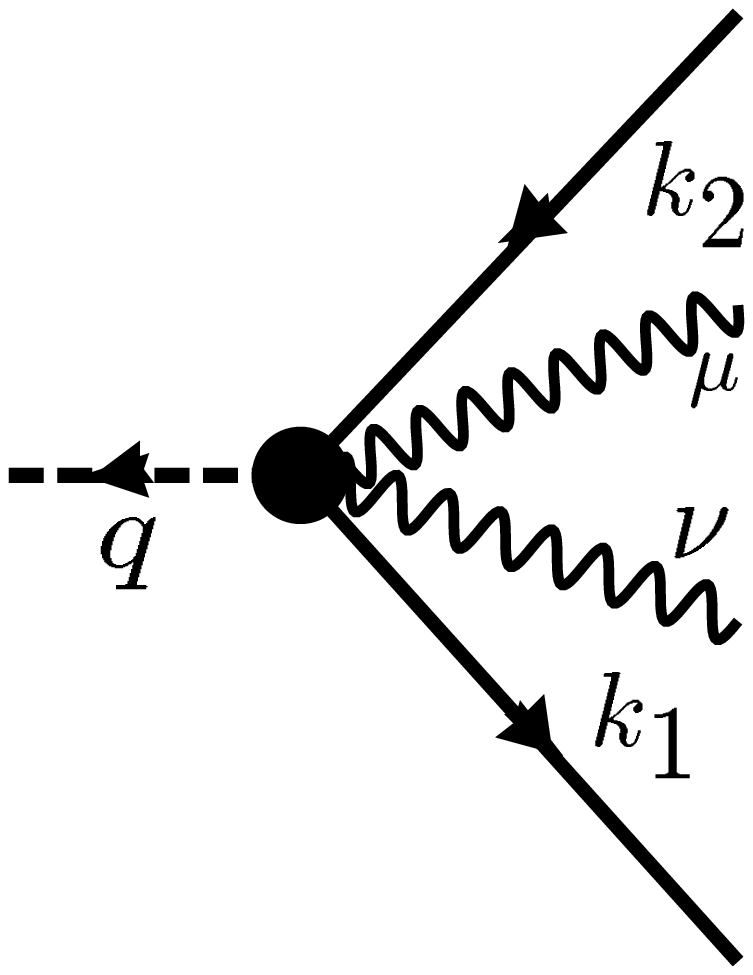,
                 scale=0.25}
    \put(-270.0,0.0){\makebox(1.0,0.0){
      $= \; g^2 \; a \; \biggl\lbrack 
          (r-s) \; c_{q+k_1+k_2,\mu}
              + (r+s) \; c_{q-k_1-k_2,\mu} \biggr\rbrack 
        \delta_{\mu\nu} \gamma_5 \sigma_\alpha $}}
      \end{picture}
   \\
   && \nonumber
\eea
where
\be
   c_{k,\mu} = \cos\frac{k_\mu a}{2} \; , \;
   \widehat{k}_\mu = \frac{2}{a} \sin\frac{k_\mu a}{2} .
\ee

\end{appendix}


%
%



\end{document}